# Local tuning of WS$_2$ photoluminescence using polymeric micro-actuators in a monolithic van der Waals heterostructure

*Francesco Colangelo*[*,1], *Andrea Morandi*[1,4,†], *Stiven Forti*[2], *Filippo Fabbri*[1,2,3], *Camilla Coletti*[2,3], *Flavia Viola Di Girolamo*[4,††], *Alberto Di Lieto*[4], *Mauro Tonelli*[4], *Alessandro Tredicucci*[1,4,5], *Alessandro Pitanti*[**,1] *and Stefano Roddaro*[1,4]

[1] NEST, Scuola Normale Superiore and Istituto Nanoscienze-CNR, Piazza San Silvestro 12, I-56127 Pisa, Italy.

[2] Center for Nanotechnology Innovation @NEST, Istituto Italiano di Tecnologia, Piazza San Silvestro 12, I-56127 Pisa, Italy.

[3] Graphene Labs, Istituto Italiano di Tecnologia, Via Morego 30, I-16163 Genova, Italy.

[4] Department of Physics "E. Fermi", University of Pisa, Largo Bruno Pontecorvo 3, I-56127 Pisa, Italy.

[5] Fondazione Bruno Kessler (FBK), Via Sommarive 18, 38123 Povo, Trento, Italy.

**Corresponding authors:** [*]francesco.colangelo@sns.it**,** [**]alessandro.pitanti@sns.it

**Abstract**

The control of the local strain profile in 2D materials offers an invaluable tool for tailoring the electronic and photonic properties of solid-state devices. In this paper, we demonstrate a local engineering of the exciton photoluminescence (PL) energy of monolayer tungsten disulfide (WS$_2$) by means of strain. We apply a local uniaxial stress to WS$_2$ by exploiting electron-beam patterned and actuated polymeric micrometric artificial muscles (MAMs), which we implement onto monolithic synthetic WS$_2$/graphene heterostructures. We show that MAMs are able to induce an in-plane stress to the top WS$_2$ layer of the van der Waals heterostructure and that the latter can slide on the graphene underneath with negligible friction. As a proof of concept for the local strain-induced PL shift experiments, we exploit a two-MAMs configuration in order to apply uniaxial tensile stress on well-defined micrometric regions of WS$_2$. Remarkably, our architecture does not require the adoption of fragile suspended microstructures. We observe a spatial modulation of the excitonic PL energy of the WS$_2$ monolayers under stress, which agrees with the expected strain profile and attains a maximum redshift of about 40 meV at the maximum strain intensity point. After the actuation, a time-dependent PL blueshift is observed in agreement with the viscoelastic properties of the polymeric MAMs. Our approach enables inducing local and arbitrary deformation profiles and circumvents some key limitations and technical challenges of alternative strain engineering methods requiring the 2D material transfer and production of suspended membranes.

Among the wide number of existing two-dimensional (2D) materials[1,2], transition metal dichalcogenides (TMDs)[3,4] have gained a preferential attention due to their gapped electronic structure, for their possibility of efficiently emitting light, even at room temperature[5], and for their wide possible applications when stacked in van der Waals heterostructures[6–8]. Monolayer tungsten disulfide[9] ($WS_2$) is one of the TDMs and shows a strong photoluminescence (PL) around 2 eV[10]. The PL of $WS_2$ mostly originates from the recombination of neutral excitons[11] and its energy is expected to be a function of the local strain[12–17] of the layer. Enabling of the local control of strain in $WS_2$, and in general in TMDs, can give access to the observation of many interesting physical phenomena such as exciton funneling[18–20], optical anisotropy and local band gap modulation[21,22], and pseudo-magnetic fields[17,23,24]; all these phenomena can be explored in a wide range of strain thanks to the remarkably good mechanical properties of monolayer $WS_2$[25]. Common methods to strain 2D materials exploit the substrate deformation[12,14,15,26–31]. These methods offer in principle a good degree of control on the strain magnitude but they do not allow a local design of the strain profile. On the other hand, alternative approaches such as the direct growth or the transfer of TMDs onto non-flat substrates emerged as promising routes to achieve strongly localized strain puddles[19,21,32–34]. However these methods lack control on the induced strain. The use of suspended 2D materials structures has been demonstrated to be an effective way to design non-homogeneous, and even anisotropic strain profiles on micrometric scales, either by using a pressure load[35–37] or micro-electro mechanical strain platforms[38–40]. But additional difficulties arise when working with suspended 2D layers due to the fragility of devices and to the formation of random strain puddles during the nanofabrication[41]. Similar limitations hold also for nano-indented suspended membranes that, while they can be an intriguing approach to study the effect of strain at the nanoscale[18,42,43], at present they have been solely used to characterize the mechanical properties of 2D materials[25,41,44]. As a consequence, inhomogeneous and anisotropic strain profiles have been poorly explored in most of 2D materials and $WS_2$ monolayers have been strained only by employing bendable/stretchable substrates[12,14,15,31] or substrate thermal expansion[16], thus only attempting homogeneous strain configurations.

Nevertheless, an appealing alternative strain engineering technique has been recently developed by some of us and it has demonstrated to be suitable to generate a wide range of controlled, spatially inhomogeneous and anisotropic strain fields in suspended graphene[45], but in principle it can be applied to other 2D materials as well. Indeed, the technique exploits patternable polymeric micrometric actuators, or micrometric artificial muscles (MAMs), made of poly-methyl-methacrylate (PMMA), that contract when irradiated by a sufficiently high dose of electrons. Here, we combine MAMs with the low-friction properties of $WS_2$ grown on top of graphene on silicon carbide (SiC)[46,47] and thus we realize localized and anisotropic in-plane strain fields in monolayer $WS_2$ in a supported configuration. We demonstrate a strain-engineered spatially-modulated redshift of the $WS_2$ PL energy, with a maximum energy shift of about 40 meV. Moreover, we show that the preparation of an initial strain state is followed by a PL shift as a function of time due to the viscoelasticity property of the employed MAMs. This relaxation allows to unequivocally identifying strain as the source of the PL shift and gives access to intermediated states at lower strain in a continuous way. The described approach is based on a monolithic synthetic van der Waals heterostructures, it is scalable[48] and does not require any transfer procedure[49]. Moreover, it allows defining multiple devices with different and custom strain profiles on the very same sample and it is expected to work with most of 2D materials.

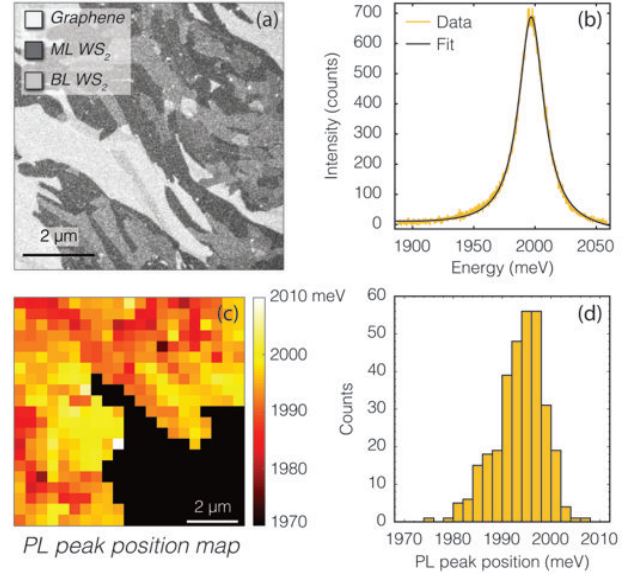

**FIG. 1.** Characterization of the as-grown sample. (a) The presence of $WS_2$ layers can be identified by scanning electron microscopy: graphene (bright), monolayer $WS_2$ (ML, dark gray) and bilayer $WS_2$ (BL, light gray). (b) Typical PL spectrum obtained on the monolayer (ML) $WS_2$. (c) Characteristic spatial distribution of the PL peak position. No PL signal is observed in the black pixels, consistently with the absence of $WS_2$ layers. (d) Histogram of the entries of panel (c), the PL energy range goes from 1980 to 2000 meV.

The fabrication of the sample starts with the growth of graphene on 6H-SiC(0001) chips by thermal decomposition (see Supporting Information). This well-established procedure can be used to obtain very large areas (centimeter-size regions) of single as well as multi-layer graphene. In our case we grow two graphene layers, i.e., a bilayer of graphene (BLG) on SiC to improve the overall flatness of the sample. In a final growth step, $WS_2$ is grown on top of the BLG by chemical vapor deposition (CVD), starting from S and $WO_3$ powders as solid precursors (see Supporting Information for further details). From the resulting shape of the CVD grown material it can be assessed that $WS_2$ layer consists of a multigrain or poly-crystalline film of $WS_2$ which covers a large portion of the BLG surface (about 50%) and mostly contains regions with one or two monolayers, as visible in the representative atomic force microscopy (AFM) map reported in the

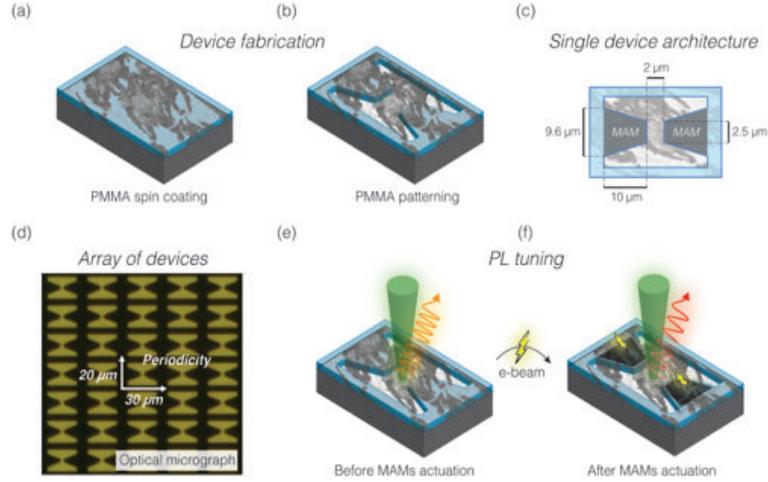

**FIG. 2.** Device fabrication steps: (a) The as-grown sample is spin-coated in PMMA; (b) selected regions are revealed by e-beam lithography and alignment markers are defined. (c) Single device architecture details. The two trapezoidal PMMA regions in gray are the one used as MAMs. (d) Optical micrograph of a portion of defied devices. In black is visible the remaining PMMA. After fabrication, the initial PL emission spectrum (e) is compared with the one obtained (f) after the actuation of the MAMs.

Supplementary Materials. Line-profiles of AFM topographic data have been used to identify the different materials from their relative heights. Figure 1(a) shows as the different materials can also be identified by considering contrast levels in scanning electron microscopy (SEM) images by setting the e-beam energy at 2 keV. The material identification is obtained from a cross-correlation with PL and AFM maps acquired in the same positions (see Supplementary Materials). SEM images show that the largest contrast steps are due to the $WS_2$ layers, as indicated in the figure. Weaker contrast steps are also visible in correspondence of BLG/SiC terraces. The PL of the sample is studied employing an *inVia Renishaw* system working at a low power of 0.5 mW and collecting the signal with a x100 objective with 0.85 NA. As-grown $WS_2$ shows a strong room-temperature PL signal when excited with 532 nm laser light. We fitted the light emission of $WS_2$ with a single Voigt lineshape (see Fig. 1(b)), which takes into account inhomogeneous broadening coming from atomic defects, grain boundaries and the finite laser spot size (~500 nm). The spatial PL distribution was studied by performing a set of micro-PL maps in different positions of the sample. We characterize the PL signal by fitting the spectrum of each point in the maps and we observe an average small spatial PL inhomogeneity, mainly due to the partial coverage of monolayers $WS_2$ and the presence of nanometric adlayers islands, which are expected to affect the emitted light properties[10,50]. The 10.5 μm × 10.5 μm spatial map of the PL peak position of Fig. 1(c) is representative of the PL properties of our van der Waals heterostacks: no relevant emission can be observed in regions where only graphene is present (black pixels); the remaining regions show an emission peaks of variable intensity and centered in the energy range going from 1980 to 2000 meV, as clear from the corresponding histogram in Fig. 1(d). In summary, our sample consists of continuous BLG with on top micrometric areas of monolayer $WS_2$ which form the van der Waals heterostructures and which show good PL properties. Although the $WS_2$ regions are randomly distributed and partially cover the sample surface, they can be quickly individuated using SEM imaging and exploited for strain-engineering experiments.

The two-MAMs device architecture employed for the strain engineering experiments is sketched in Fig. 2. After the preliminary characterization, the sample is spin-coated with PMMA (AR-P 679.04) for 60 seconds at 4000 rpm and then soft-baked for 15 minutes at 90 °C to evaporate the resist solvent. As a result, a uniform 110 nm thick layer of PMMA covers the whole sample surface (Fig. 2(a)). Then, we use standard e-beam lithography to pattern into the PMMA layer a set of "access windows" revealing portions of the $WS_2$ top surface, and alignment markers, see Fig. 2(b). Each access window identifies a single device. The two horizontal trapezoidal regions of PMMA implement the two opposite pulling MAMs that will contract when exposed to a high dose of electrons. It is worth to note that such a design ensures that one of the two ends of each MAM is embedded – and thus anchored – to the external PMMA frame while the opposite MAM tips are free from the mechanical load of PMMA. In this way, even if the two MAMs are supported by a superlubric layer, they can still apply a tensile stress to the material in between them. In addition, the PMMA removal allows SEM and AFM imaging of the sample on the strain-engineered regions. The nominal distance between the tips of the two MAMs is 2 μm and corresponds to the typical size of the monolayer $WS_2$ domains of our sample. A detailed scheme depicting the architecture of devices is reported in Fig. 2(c). It is worth to underline that this approach is easily scalable and many devices can be prepared in a single lithographic step, as shown in the optical micrograph of Fig. 2(d) which shows a portion of the batch of the 600 fabricated devices. A set of 20 devices was used for $WS_2$ pulling experiments while many others were exploited for explorative tests and calibrations; a quick device screening is possible thanks to the compatibility of the technique with *in situ* SEM imaging and allows selecting devices with a continuous layer of $WS_2$ between the pair of MAMs for strain engineering experiments. The actuation of the MAMs is obtained by exposing both the trapezoidal PMMA regions with equal doses and using an e-beam energy of 2 kV. By contracting, the MAMs are expected to induce a specific strain pattern in the $WS_2$ layer and therefore to write a specific spatial pattern of the $WS_2$ PL energy. Indeed, as mentioned before, the exciton PL depends on the local strain. In particular the PL energy is expected to be proportional to the volumetric component of the strain[12,14–16] (i.e. to the trace of the tensor strain) and mainly due to a change of the material band gap[14,17,24]. Since the total exposure dose determines the final PMMA shrinkage magnitude, the pulling stress of MAMs and, as a consequence, the magnitude of the induced PL shift can be tuned. In order to have differential measurements of the device properties, SEM, AFM and PL maps were measured before and after the MAMs actuation as sketched in Fig. 2(e-f).

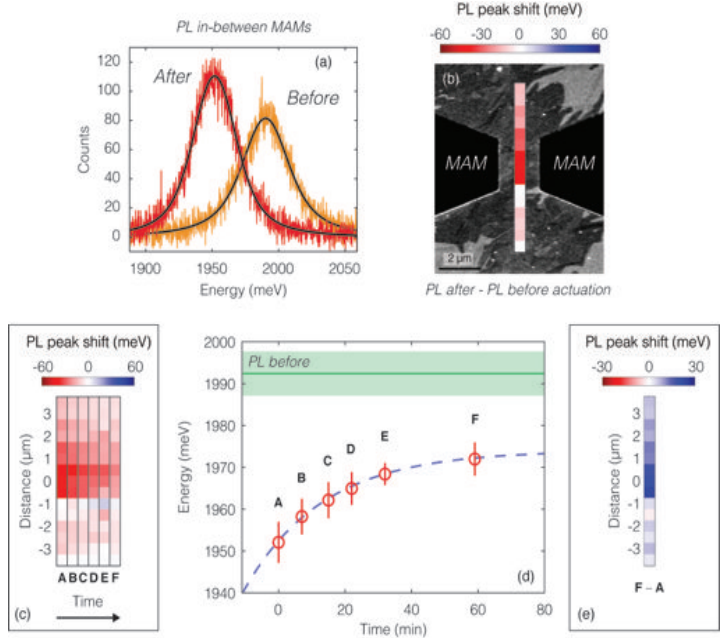

**FIG. 3**. (a) Comparison of two PL spectra acquired at the center of one of the studied device before and after the MAMs actuation. Black lines are best Voigt fits and indicate a PL redshift of about 40 meV. (b) SEM micrograph of the device with a color plot of the PL shift profile in overlay. The reference PL peak position corresponds to the one before the MAMs actuation. (c) Consecutive PL shift profiles acquired at different times. (d) Time evolution of the PL peak position (red circles) as the center of the device. The green line indicates the PL energy before the MAMs actuation, while the blue dashed line is an exponential best fit. (e) PL shift between the first measurement after the MAMs actuation (scan "A") and the last one (scan "F").

It is worth to stress that the discussed experimental method crucially depends on the ability of $WS_2$ to slide on top of the graphene layer. Although there are experimental evidences that sub-micrometric single crystals of $WS_2$ can slide on top of graphene with negligible friction[46,47], it is not clear if such a superlubric property holds also for micrometric portions of multigrain $WS_2$ layers. In order to shed light on the tribological properties of our $WS_2$/BLG heterostructure interface, we tested devices showing a discontinuous $WS_2$ layer between the two MAMs and we pulled and dragged the $WS_2$ layer for hundreds of nanometers on the BLG underneath. This is possible by exploiting the high shrinkage magnitude (>5%) of MAMs when irradiated with doses of hundreds of $mC/cm^2$. Results of our sliding tests (see Supplementary Materials) allows us to conclude that both the static and dynamic friction do not play any significant role in our experiments.

The PL properties of $WS_2$ under stress are investigated by performing a set of micro-PL line scans across the gap between MAMs connected with a continuous layer of $WS_2$. We used a low pump power of 0.5 mW and 1 second of integration time aiming to obtain a good signal-to-noise ratio while minimizing the local heating effect, which is known to influence the viscoelastic relaxation of MAMs[45]. In addition, these experimental conditions allow reducing the total acquisition time for a single PL map on time scales much shorter than the typical relaxation time of MAMs. The step size of the line scans is of 500 nm, which approximately corresponds to the laser spot size. A lag time typically ranging from 10 to 15 minutes exists between actuation and the start of PL measurements. Figure 3 shows the results of a strain engineering experiment performed on one of the devices. A comparison of PL spectra collected on the central region of the device before and after the MAMs exposure is reported in Fig. 3(a). Using an actuation dose of 15 $mC/cm^2$, we observe a redshift of about 40 meV (~12 nm in wavelength) of the $WS_2$ light emission spectrum. Moving away from the central region of the device the redshift decreases and its spatial modulation is depicted in Fig. 3(b) where a color map of the PL shift overlays the SEM micrograph of the device. The plotted PL shift values are obtained by subtracting pixel-by-pixel the PL energy before and after the actuation. The PL redshift pattern in between the two MAMs can be understood as a consequence of the inhomogeneous MAMs-induced strain profile. Indeed, based on our numerical simulations (see Supplementary Materials) and on what already observed in suspended graphene[45], the strain pattern is expected to be anisotropic, tensile, and strongly localized in between the two MAMs. We note also that the PL shift profile does not exactly match the mirror symmetry of the device architecture but the discrepancy can be understood based on the non-trivial and asymmetric shape of the studied $WS_2$ flake. It is worth to mention that during strain-engineering experiments, *in situ* monitoring of devices revealed a systematic mechanical breakdown of the $WS_2$ layers by employing doses exceeding 30 mC/cm$^2$, limiting the possibility to further shift the PL energy in our multigrain sample.

Strain engineering experiments that used MAMs on suspended graphene clearly showed that the induced strain in the 2D layer relaxes as a function of time due to the viscoelastic relaxation phenomena in the MAMs[45]. In order to investigate this aspect we monitored the time evolution of the PL signal coming from the cross-section of the device during an hour. The scan lines at different times reported in Fig. 3(c) show that we consistently observe a similar relaxation also in the current experiment. In Fig. 3(d) a plot of the PL energy averaged on the three central pixels as a function of time shows that the decay of the PL strain-induced redshift nicely follow an exponential law (blue dashed line) with a characteristic time of about 25 min, in agreement with what expected for a viscoelastic relaxation behavior. Moreover, we note that after the MAMs actuation, the PL position does not recover its original value (green line in Fig. 3(d)) and, in addition, the subsequent PL blueshift mainly involves regions where a larger stress intensity is expected. Indeed, by considering the relative shift between the first scan line measurement after the MAMs actuation ("A") and the last one ("F"), the PL blueshift largely occurs only the central region of the device (see Fig. 3(e)). All these facts not only further confirm the origin of the observed shift but also indicate the possibility to use MAMs to continuously tune the strain in a wide range of intensity.

In order to have a rough but independent estimation on the deformation induced in the $WS_2$ layer we performed numerical simulations using COMSOL Multiphysics. As described in the Supplementary Materials, we modeled a whole device emulating the main characteristics of real devises (mechanical properties, dimensions, clamping conditions, frictional properties). In addition we simulated the MAMs actuation by adding an initial negative and isotropic pre-strain in the MAMs domains and eventually we let the system relax to its mechanical equilibrium. By considering a dose of 15 mC/cm$^2$ and our shrinkage-per-dose calibration (obtained testing devices containing discontinuous $WS_2$), numerical results indicate a volumetric component of the strain of about $\varepsilon_{xx} + \varepsilon_{yy} = 0.54\%$ at the central point of the device. As a consequence, our PL shift rate estimation is of about -74 meV/%strain, with an uncertainty of about 40% mainly due to the shrinkage calibration. This gauge value is in good agreement with theoretical predictions[15,17,24] and with two out of the three available experimental works[15,16] reporting similar experiments, while it results significantly larger if compared to the remaining one[14]. The latter discrepancy might be due to a subtle but significant non-ideal strain transfer efficiency that can exist in experiments that use deformable substrates[12,16,28].

In conclusion, we studied the PL properties of a monolayer $WS_2$ stacked on top of BLG and subject to a strongly localized uniaxial tensile stress. We created such a stress configuration by combining micrometric artificial muscles with the ability of $WS_2$ to slide on graphene. The local PL energy is directly connected to the local $WS_2$ energy band gap that we pattern with micrometric resolution by mean of the induced strain profile. We obtained a maximum PL redshift of about 40 meV at the position where the top strain is expected. After the preparation of the initial MAMs-induced PL redshift profile, we observed a partial and gradual tuning of the initial PL energy as a function of time, as previously reported in similar strain engineering experiments using PMMA-based MAMs. The used patternable polymeric actuators can be designed with great freedom and implemented by conventional e-beam lithography accessing a great variety of strain fields. In addition, in the present work we demonstrated an implementation of MAMs that does not require the creation of fragile suspended microstructures nor the transfer of $WS_2$ from its growth substrate. The combination of MAMs technology with the small frictional force occurring between 2D materials bound by van der Waals interactions opens the route to advanced strain-engineered devices.

SUPPLEMENRARY MATERIALS

See the supplementary material for additional information concerning sliding and adhesion properties of the $WS_2$ layer, the growth of the sample and the numerical simulation of devices.

AUTHOR INFORMATION

**Corresponding Authors**

francesco.colangelo@sns.it, alessandro.pitanti@sns.it

**Present Addresses**

† Optical Nanomaterial Group, Institute for Quantum Electronics, Department of Physics, ETH Zürich, Auguste-Piccard- Hof 1, 8093 Zürich, Switzerland.

†† Istituto Nazionale di Fisica Nucleare, L.go Bruno Pontecorvo 3, 56127, Pisa, Italy.

**Author Contributions**
The manuscript was written through contributions of all the authors. F.C. and A.M. equally contributed to this work.

**Acknowledgment**

The research leading to these results has received funding from the European Union's Horizon 2020 research and innovation program under grant agreement No. 785219 – GrapheneCore2. F.C. and S.R. acknowledge the financial support from the project QUANTRA, funded by the Italian Ministry of Foreign Affairs and International Cooperation. S.R. acknowledges the support of the PRIN project Quantum2D funded by the Italian Ministry for University and Research.

# Supplementary Information

*Francesco Colangelo*[*,1], *Andrea Morandi*[1,4,†], *Stiven Forti*[2], *Filippo Fabbri*[1,2,3], *Camilla Coletti*[2,3], *Flavia Viola Di Girolamo*[4,††], *Alberto Di Lieto*[4], *Mauro Tonelli*[4], *Alessandro Tredicucci*[1,4,5], *Alessandro Pitanti*[**,1] *and Stefano Roddaro*[1,4]

[1] NEST, Scuola Normale Superiore and Istituto Nanoscienze-CNR, Piazza San Silvestro 12, I-56127 Pisa, Italy.
[2] Center for Nanotechnology Innovation @NEST, Istituto Italiano di Tecnologia, Piazza San Silvestro 12, I-56127 Pisa, Italy.
[3] Graphene Labs, Istituto Italiano di Tecnologia, Via Morego 30, I-16163 Genova, Italy.
[4] Department of Physics "E. Fermi", University of Pisa, Largo Bruno Pontecorvo 3, I-56127 Pisa, Italy.
[5] Fondazione Bruno Kessler (FBK), Via Sommarive 18, 38123 Povo, Trento, Italy.

**Corresponding authors:** [*]francesco.colangelo@sns.it, [**]alessandro.pitanti@sns.it

**Keywords:** *$WS_2$, strain engineering, photoluminescence, van der Waals heterostructures, polymeric actuators, MAM*

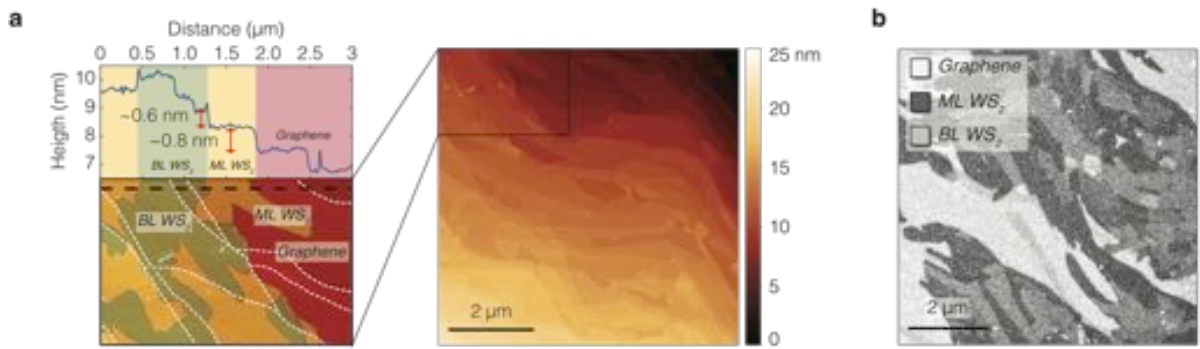

**FIG. SA.** Topographic characterization of the as-grown sample. (a) Topographic AFM image of the monolithic $WS_2$/BLG/SiC sample. Graphene and $WS_2$ regions can be identified in the cross-section on the left side of the panel. (b) $WS_2$ can also be identified by scanning electron microscopy due to the different contrast of graphene (bright), monolayer $WS_2$ (ML, dark gray) and bilayer $WS_2$ (BL, light gray).

# 1. Sliding and adhesion phenomena between PMMA, graphene and WS$_2$

Data reported in Fig. S0 offer an insight on the role of friction in our experiments: we report two AFM pictures before (Fig. S0(a)) and after (Fig. S0(b)) the actuation of two MAMs patterned according to the sketch in Fig. 2(c) in the main manuscript. In Fig. S0(a) and S0(b), the MAMs are visible on the left and right ends of the scan in a white color. At the center of the scan in Fig. 3 S0(a), a large number of small PMMA residues cover the very top surface of the sample between the two MAMs. While these residues can be easily minimized and their presence is typically undesirable, here they allowed us to estimate the local displacement of the top surface and to tracking their individual positions before and after the actuation. In the figure, the top layer is identified as WS$_2$, based on AFM data and on a local mapping of the PL. We notice that the layer is mostly composed by monolayer WS$_2$, with limited bilayer areas. It is worth noting that the two MAMs are located on top of two mechanically independent regions of WS$_2$ monolayers, which are indeed divided by a crack indicated in Fig. S0(a) by the two black arrows. In this sliding experiment the MAMs are actuated by an e-beam exposure with a large dose of 200 mC/cm$^2$ so that the resulting significant displacement of the two WS$_2$ regions allows better assessing the tribological properties of the van der Waals heterostack. The shrinkage resulting from such a dose is visible in panel (b) where the distance between the MAMs grows significantly: a 500 nm wide fracture opens between the two decoupled regions of WS$_2$ (black arrows in Fig. S0(b)). The emerging strip is free from polymeric residues and clearly originates from the sliding of the top WS$_2$ layers; this process in fact uncovers a portion of the underlying graphene, which is atomically clean since it was protected by the WS$_2$ layer during the PMMA spin-coating (see also the trhee-dimensional rendering in Fig. S0(c)). The topographic discontinuity between the clean strip and the covering top regions confirms the presence of a monolayer WS$_2$/graphene stack.

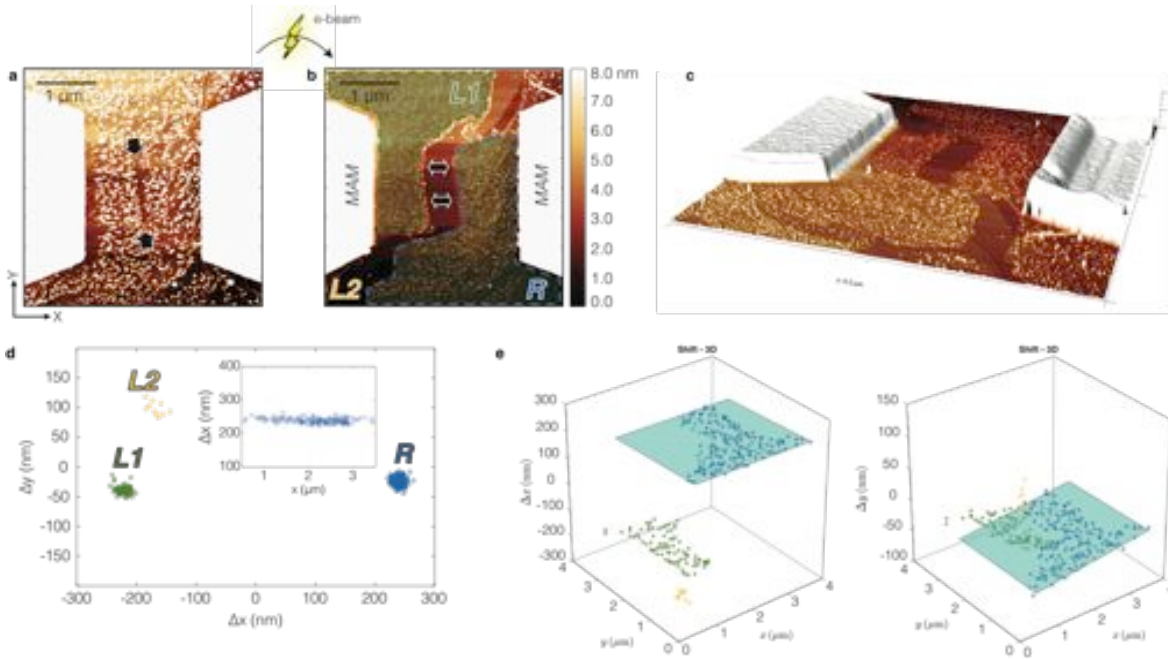

**FIG. S0**. MAMs-induced displacement of WS$_2$. (a) Atomic force micrograph of one of the studied devices. Two polymeric regions (MAMs) are visible in white (height > 8 nm) on the left and right ends of the scan. WS$_2$ covers the full free sample surface in the figure, but a small fracture is visible in the center (highlighted by the arrows). Small PMMA residues are visible on top of the WS$_2$ and here serve as a reference to estimate displacements. (b) After the actuation of the MAMs, WS$_2$ breaks in two main parts and slides (mostly) sideways. A residue-free region emerges in the center of the scan as a vertical/diagonal fracture with a width of approximately 500 nm. (c) Three-dimensional rendering of AFM data of panel (b). (d) PMMA residues are used to estimate the local $(\Delta x, \Delta y)$ displacement of the WS$_2$ layer (see Supplementary Information). Values fall in three distinct groups corresponding to three difference regions "R", "L1" and "L2". Inset: the lack of correlation between $\Delta x$ and the position of $x$ in the "R" group indicates that – within the experimental errors – WS$_2$ undergoes a *rigid* translation. (e) Results of the fits for the displacement values as a function of the position.

As mentioned in the main text, if a sensible amount of static friction is present between WS$_2$ and the BLG, then the pulling stress of MAMs will accumulate just close the MAMs ends and: (i) the WS$_2$ is not expected to slide and/or (ii) it is expected to crack in the high-stressed region of the WS$_2$ (i.e. close the MAMs tip). Since we observe the WS$_2$ to slide we have a clear indication that static friction at the WS$_2$/BLG interface is relatively small and that we are able to overcome it. In order to estimate the impact of possible non-negligible dynamic friction we measured the average deformation of the WS$_2$ by taking advantage of the possibility to measure the $(\Delta x, \Delta y)$ displacement of a large number of PMMA residues (see also below and Fig S1 for further details). Indeed, a significant dynamic friction will result in a final *tensile* deformation of the WS$_2$. The particle tracking results are reported in panel (d): three displacement "groups" can be clearly identified and are named "R", "L1" and "L2". The groups correspond to three spatially separated "rafts" of WS$_2$: the "R" portion of the material is attached to the right MAM and moves towards the right side of the scan by about 250 nm. A large "L1" portion is connected to the opposite MAM and moves in the opposite direction. A smaller fraction "L2" is visible on the bottom left corner and displays a non-negligible vertical displacement caused by the lateral shrinking of the MAM. In the inset of panel (d) we report the horizontal displacement $\Delta x$ versus the horizontal position $x$ for points belonging to the group "R", which covers most of the bottom-right region of the scan. As visible from the plot, we observe no correlation between the displacement and the residue position, consistently with a *rigid translation* of the WS$_2$. In order to quantify the average deformation, the $\Delta x$ and $\Delta y$ displacements versus $(x, y)$ were fitted with a linear model where $\Delta x = u_{xx}x + u_{xy}y + q_x$ and $\Delta y = u_{yx}x + u_{yy}y + q_y$, respectively, and where $u_{ij}$, is the average displacement gradient tensor and $(q_x, q_y)$ is the average displacement (see Fig. S0(d)). After eliminating rotation components by symmetrizing the $u_{ij}$ tensor, the averaged strain tensor $\varepsilon_{ij}$ can be obtained. As a result, most of the WS$_2$ regions show a deformation compatible with zero frictional forces, for example the "R" group shows a hydrostatic strain component $\bar{\varepsilon}(\%) = \sum \varepsilon_{ii}/2 = -0.05 \pm 0.05$. On the other hand, it is worth to note that there is a relevant exception. The deformation resulting from the tracking of particles located in the upper right region (just above the right MAM) indicates a hydrostatic strain of about -24%. Such a high level of compression is typically relaxed in suspended 2D materials by out-of-plane displacements of the layer. In our case, we also noted the presence of a wrinkle, which is visible in the same WS$_2$ region after the actuation, and provides a further evidence of the very low mechanical interaction between WS$_2$ and graphene. Our sliding tests and particle tracking analysis allows us to conclude that both the static and dynamic friction do not play any significant role in our experiments.

The analysis of the WS$_2$ displacement reported in Fig. S0 has been obtained by tracking the movement of individual PMMA residues in a device with non-ideal cleaning. The tracking is based on the two AFM scans reported also in Fig. S1. As visible in the two magnified views of panel (a), the pattern of the PMMA residues is very often clearly recognizable before and after the actuation. The movement of individual points can thus be tracked with nanometric precision. A set of 285 individual point pairs were manually collected and the resulting displacements are summarized in Fig. S0. The points can be clearly separated in three distinct groups which are identified by name and color: the right

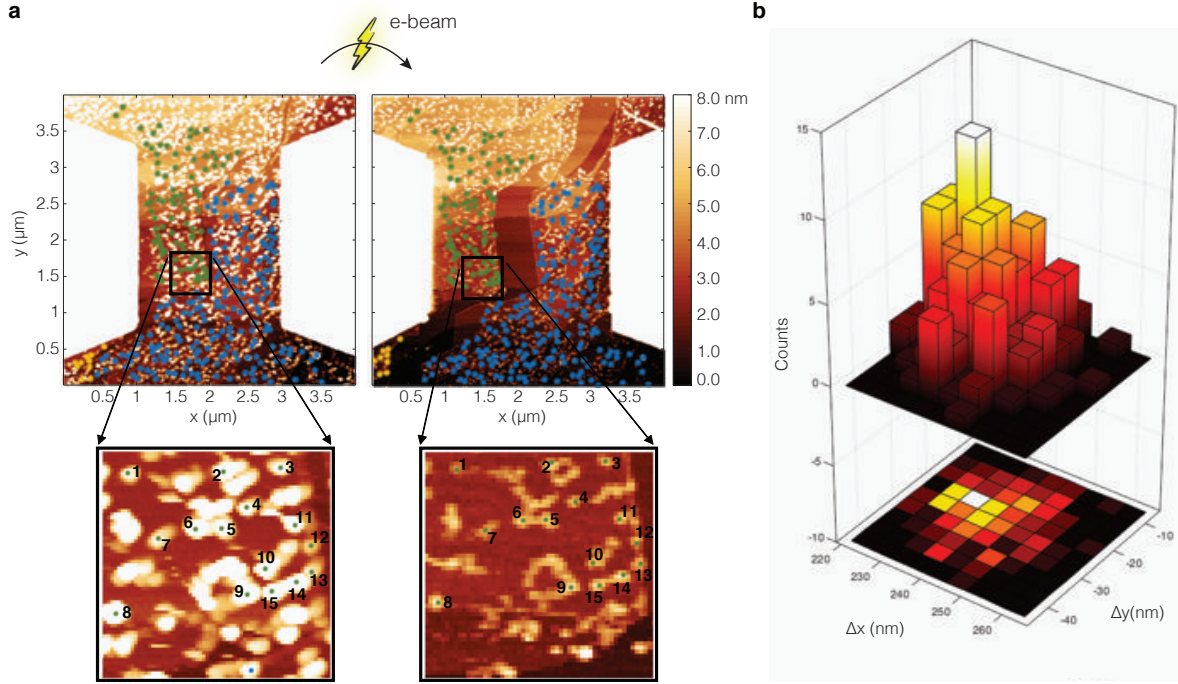

**FIG S1.** (a) AFM scans before (left) and after (rigth) the actuation of the MAMs whose tip are visible in white. The different displacement groups are highlighted in different colors as an overlay to the topographic scan. Magnified views below the panels shot the correspondence between the residues in the two scans. Displacements can be measured with a precision of about 10 nm (panel (b)).

"R" group in blue (188 points); the first left "L1" group in green (87 points); the second left "L2" group in orange (10 points). The dispersion of the points is of the order of 10 nm (see histogram and colorplots in Fig. S1(b)) and appears to be dominated by the measurement error, which is due to the lateral size of the PMMA residues and to the small variations between the two scans. In fact, the first scan appears to have partially "cleaned" the top surface and has led to a significant reduction of their size. Nevertheless, PMMA patterns are in most of cases still easy to identify.

Another key ingredient of the experiment is constituted by the *adhesion* between the MAMs and the two-dimensional layers of the $WS_2$/graphene/SiC stack. In this case, revealing studied can be performed directly starting from scanning electron micrographs collected before and after the MAMs actuation. Figure S2 display an RGB composite picture where the post-actuation image was used to set the R (red) channel while the pre-actuation image was used to set the remaining B (blue) and G (green) channels. Since the PMMA appears as darker (black, with the brightness/contrast settings we used) than the $WS_2$/graphene/SiC, a *contraction* of the MAM corresponds to the emergence of a *red* region at the edge of the MAMs. Different behaviors can be easily spotted in the figure and have been highlighted by numbers in overlay: in regions 3, 5 and 6 we observed a significant contraction; in region 1 we have a slight contraction; in regions 2 and 4 no evidence of contraction is observed. The visual information delivered by the composite image can be analyzed in a quantitative way by performing contrast averages along the MAM's edges in the regions cited above. In particular, vertical pixel lines were aligned using a threshold on the pre-actuation image and averaged in each region. The vertical position (abscissa in the plots) was then finally rescaled by a factor $\cos(30°)$ to take into account the inclination of the MAM's edge and obtain an estimate of the *perpendicular* displacement. The results are reported in a set of subpanels and confirm the information encoded in the color: regions 3, 5 and 6 contract by about 100 nm; region 1 contracts by about 40 nm; no visible contraction is reported in regions 2 and 4.

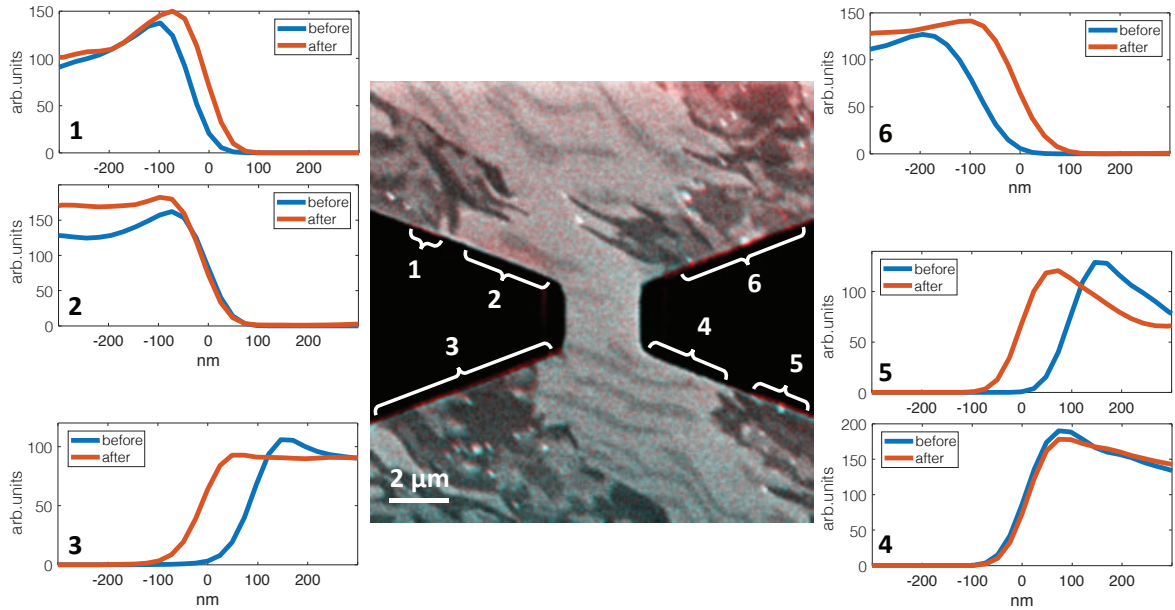

**FIG S2.** Main panel: composite RGB image combining a SEM picture after (red channel) and before (green and blue channel) the actuation of the MAMs. The movement of the edge of the MAMs can be identified in terms of colors in the scan. A set of cross-sectional contrast profiles were extracted in the specific regions in overlay and reported in the corresponding subpanels.

It is interesting to note that the regions were the MAMs contract are in almost perfect agreement with the presence of $WS_2$ at the edge of the polymeric region. Differently, when the MAMs are directly on top of graphene no contraction is observed. The obvious interpretation of this behavior is that the PMMA displays a good adhesion to both graphene and $WS_2$. The easy sliding of the discontinuous portions of $WS_2$ on top of graphene (see previous AFM study) thus allows the MAMs to contract by a large amount. When a continuous $WS_2$ layer is placed between the MAMs, as in the case of graphene or in the pulling experiment on $WS_2$ reported in the main text, the MAMs can only contract to the extent to which the two-dimensional material is able to stretch. We note that for a sufficiently large exposure dose the MAMs were found able to tear apart both $WS_2$ and even the underlying graphene layers. It is unclear whether the fracture is nucleated by defects in the material or to the reaching of the intrinsic yield strength of the two materials.

## 2. Heterostructures growth

The graphene samples are grown on commercial semiconducting 6H-SiC(0001) single-side polished wafers, from *SiCrystal GmbH*. The wafers were on-axis with a declared unintentional miscut angle below 0.5° and an extrinsic n-type doping achieved via N implantation of about $10^{18}$ cm$^{-3}$. The SiC substrates were hydrogen etched to remove chemomechanical scratches and to regularize the surface into a step-terrace structure. The clean and atomically flat Si-terminated surface is then graphitized following the recipe introduced in 2009 by Emtsev and coworkers [1], i.e. by annealing the sample with a background Ar atmosphere at 1395 °C for 8 minutes. In this way, we obtain a graphene bilayer on the SiC C-rich $(6\sqrt{3}\times6\sqrt{3})R30$ reconstruction [1-5]. The whole process is carried out in a commercial *Black Magic*™ from *Aixtron*. Figure S3(a) shows the Raman spectrum of graphene grown on SiC substrate. The full width half maximum of the 2D mode, ~ 45 cm$^{-1}$, and its composite lineshape indicates the presence of bilayer graphene. Moreover, the clear presence of the G mode and the 2D/G intensity ratio confirms the growth of bilayer graphene.

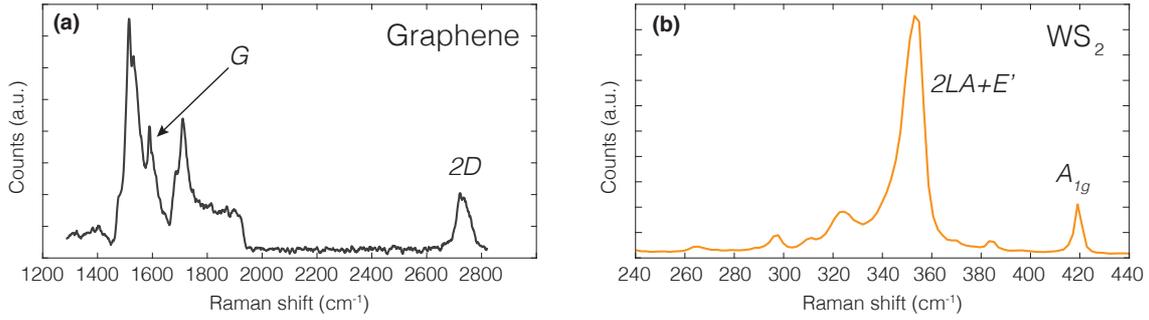

**FIG S3**. (a) Raman characterization of graphene growth on SiC substrate. The presence of the G peak and a composite 2D peak (around 2720 cm$^{-1}$) confirms the growth of bilayer graphene. (b) The Raman spectrum after the growth of WS$_2$ shows the typical features of crystalline WS$_2$.

The WS$_2$ growth is carried out in a standard hot-wall reactor by chemical vapor deposition (CVD) [6-7]. The precursors, employed in the synthesis, are WO$_3$ and S powders with a ratio of 1:100 in weight. The WO$_3$ powder is placed close to the substrate and it is heated up to 920 °C. The sulfur powder is heated up to 130 °C to induce the evaporation of the precursor. Argon is employed as carrier gas with a flow of 8 sccm, while the pressure in the reactor was kept at 0.5 mbar. A representative Raman spectrum of the resulting WS$_2$ is reported in Fig. S3(b). The fingerprints of the presence of highly crystalline WS$_2$ are the 2LA peak at ~ 352 cm$^{-1}$ and the A$_{1g}$ mode at about 419 cm$^{-1}$. The ratio of the 2LA and A$_{1g}$ intensities suggests the synthesis of monolayer WS$_2$ [8].

### 3. Mechanical simulation of the device

In order to estimate the strain applied by the MAMs we measured the PMMA horizontal shrinkage using the SEM after consecutive exposure steps in devices that do not present the mechanical resistance of the WS$_2$ layer and we obtained an estimated shrinkage per dose factor of 0.040 ± 0.015 %/(mC/cm$^2$) in the range from 10 mC/cm$^2$ to 50 mC/cm$^2$.. According to this calibration we have a -0.6% MAM shrinkage for a dose of 15 mC/cm$^2$. Therefore, this shrinkage value has been used as a MAM pre-strain for estimating the strain induced in the WS$_2$ by simulating the device mechanics. In order to do that, we employed COMSOL Multiphysics modeling the three-dimensional device architecture using the same dimensions we experimentally implemented. The resulting simulated geometry is reported in Fig. S4(a). The WS$_2$ has been modeled as a 0.65 nm thick and continuous layer with a 270 GPa Young's modulus of and 0.22 Poisson ratio [9]. The PMMA regions are 110 nm thick with a Young's modulus and a Poisson ration of 3 GPa and 0.4, respectively. As initial condition, the MAMs region have a -0.6% of pre-strain, the outer vertical boundaries of the device are fixed and the WS$_2$ is free to move in the horizontal direction while it is clamped to the PMMA. Once the system reaches the mechanical equilibrium, the strain tensor in the WS$_2$ layer is evaluated. The pre-strain of the MAMs clearly results in a pulling stress for the WS$_2$ layer and as a consequence induces a specific strain profile. Figure S4b show the spatial distribution of the strain tensor trace, while panel c and d show the strain component along and orthogonally the pulling axis of the MAMs, respectively. In particular, at the center of the device we found in the WS$_2$ layer a tensile strain component of about 1.02% along the pulling axis ($\varepsilon_{xx}$) and a compressive component of about -0.48% along the orthogonal direction ($\varepsilon_{yy}$) with a resulting strain anisotropy (or deviatoric component) of $\varepsilon_{xx} - \varepsilon_{yy} = 1.5\%$, while the resulting volumetric strain or the strain tensor trace is of $\varepsilon_{xx} + \varepsilon_{yy} = 0.54\%$ as mentioned in the main text.

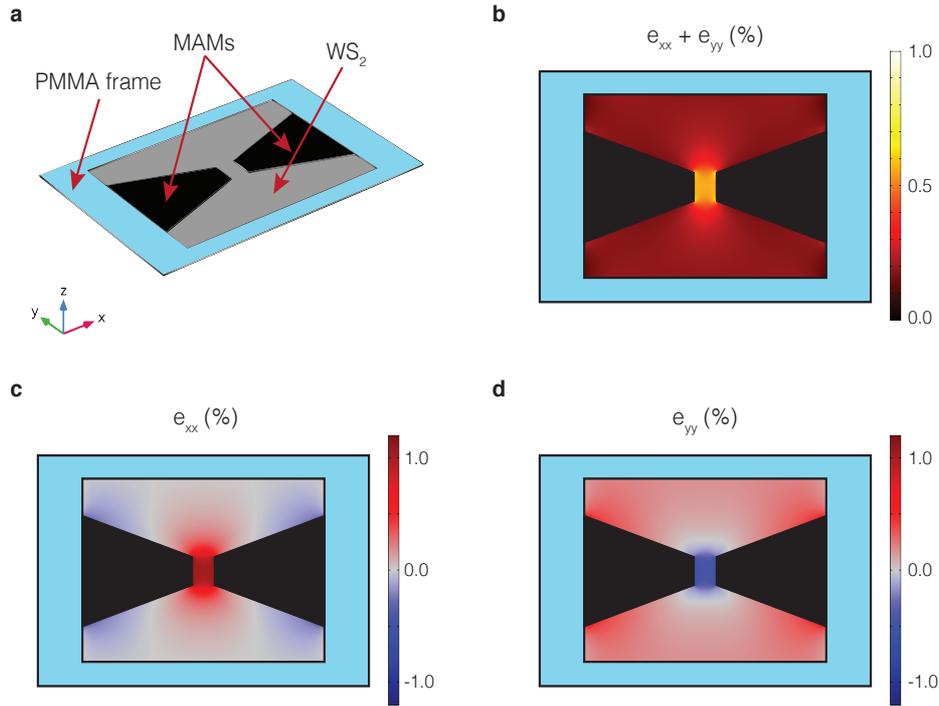

**FIG S4**. (a) Device architecture. In cyan the PMMA frame, in black the MAMs and in gray the WS$_2$ layer underneath. After that the mechanical equilibrium is reached the strain components relative to the WS2 can be obtained. The result is visible in (b), (c) and (d) which show the trace of the strain tensor, the strain component along the pulling axis of the MAMs and the strain component orthogonal to the pulling axis, respectively.